\documentclass[aps,prd,twocolumn,10pt,groupedaddress]{revtex4-1}
\usepackage{amssymb}
\usepackage{graphicx}
\usepackage{amsmath}
\usepackage{hyperref}
\usepackage{subfigure}
\usepackage{float}
\usepackage{color}
\usepackage{ulem}
\usepackage[utf8]{inputenc}

\begin{document}

\title{Mass bound for primordial black hole from trans-Planckian censorship conjecture}
\author{Rong-Gen Cai}
\email{cairg@itp.ac.cn}
\affiliation{CAS Key Laboratory of Theoretical Physics, Institute of Theoretical Physics, Chinese Academy of Sciences, Beijing 100190, China}
\author{Shao-Jiang Wang}
\email{schwang@cosmos.phy.tufts.edu}
\affiliation{Tufts Institute of Cosmology, Department of Physics and Astronomy, Tufts University, 574 Boston Avenue, Medford, Massachusetts 02155, USA}

\begin{abstract}
The recently proposed trans-Planckian censorship conjecture (TCC) imposes a strong constraint on the inflationary Hubble scale, of which the upper bound could be largely relaxed by considering a noninstantaneous reheating history. In this paper we will show that, if the primordial black holes (PBHs) are formed at reentry in the radiation-dominated era from the enhanced curvature perturbations at small scales, the TCC would impose a lower bound on the PBH mass $M_\mathrm{PBH}>\gamma(H_\mathrm{end}/10^9\,\mathrm{GeV})^2\,M_\odot$ regardless of the details for reheating history, where $\gamma$ is the collapse efficiency factor and $H_\mathrm{end}$ is the Hubble scale at the end of inflation. In particular, the current open window for PBHs to make up all the cold dark matter could be totally ruled out if the inflationary Hubble scale is larger than 10 TeV. For the case of PBHs formed in an early matter-dominated era, an upper mass bound is obtained.
\end{abstract}
\maketitle

\section{Introduction}\label{sec:int}

Perhaps the most astonishing insight on cosmology is that the phenomena at the largest scale, like the cosmic microwave background (CMB) \cite{Gorski:1996cf,Hinshaw:2012aka,Aghanim:2018eyx} and large-scale structure (LSS), could emerge from the phenomena  of quantum fluctuations at the smallest scale, which can be traced back to an accelerating expansion phase in the early Universe with shrinking comoving Hubble horizon $1/(aH)$ in the inflationary cosmology \cite{Brout:1977ix,Sato:1980yn,Guth:1980zm,Linde:1981mu,Albrecht:1982wi}, as well as some alternative scenarios (see, e.g., Ref.  \cite{Brandenberger:2009jq} for a review). Such an early inflationary phase, if it lasted long enough, would eventually stretch even the smallest quantum fluctuations of Planck size out of the Hubble horizon, after which they would become classical and frozen until reentry to be observed today. This raises the well-known inflationary trans-Planckian problem \cite{Martin:2000xs,Brandenberger:2000wr,Brandenberger:2012aj,Kaloper:2002cs,Easther:2002xe}, since, in a consistent theory of quantum gravity, these trans-Planckian quantum fluctuations should remain quantum so as not to jeopardize the effective field theory (EFT) treatment on inflation. This leads to the recent claim \cite{Bedroya:2019snp} of the trans-Planckian censorship conjecture (TCC) that no trans-Planckian mode should ever exit the Hubble horizon that would otherwise belong to the swampland.

The TCC puts a strong constraint on the duration of an early inflationary phase:
\begin{align}\label{eq:TCCa}
\frac{a_f}{a_i}<\frac{M_\mathrm{Pl}}{H_f},
\end{align}
where $a_{i,f}$ are the scale factors at the beginning and end of that inflating phase, and $H_{i,f}$ are the corresponding Hubble scales. Working with the approximation of a constant inflationary Hubble scale $H_i\approx H_f\approx H_\mathrm{inf}$, Eq. \eqref{eq:TCCa} could be translated into an upper bound on the inflationary e-folding number $N_\mathrm{inf}$ (the e-folding number at the end of inflation is fixed to be zero throughout the paper):
\begin{align}\label{eq:TCCN}
\mathrm{e}^{N_\mathrm{inf}}<\frac{M_\mathrm{Pl}}{H_\mathrm{inf}},
\end{align}
which serves as a stronger bound compared to an early estimation $N_\mathrm{inf}<M_\mathrm{Pl}^2/H_\mathrm{inf}^2$ from quantum gravity \cite{ArkaniHamed:2007ky}. If such an early inflationary phase is directly connected to the phase of standard big bang expansion with instantaneous reheating history, it could be quickly observed \cite{Bedroya:2019tba} from Eq. \eqref{eq:TCCN} that $N_\mathrm{inf}=46.2$, leading to a strong constraint on the inflationary Hubble scale and tensor-to-scalar ratio,
\begin{align}
&\frac{H_\mathrm{inf}}{M_\mathrm{Pl}}<\mathrm{e}^{-N_\mathrm{inf}}=8.4\times10^{-21},\label{eq:TCCH}\\
&r\equiv\frac{2}{\pi^2\mathcal{P}_\mathcal{R}}\left(\frac{H_\mathrm{inf}}{M_\mathrm{Pl}}\right)^2<6.8\times10^{-33},\label{eq:TCCr}
\end{align}
where $\mathcal{P}_\mathcal{R}\approx2.1\times10^{-9}$ is used from Planck 2018 \cite{Aghanim:2018eyx}. However, the upper bound in Eq. \eqref{eq:TCCr} is so strong for slow-roll inflation models that it would cause a severe fine-tuning of initial conditions.

Fortunately, the upper bound [Eq. \eqref{eq:TCCr}] could be largely relaxed by considering a noninstantaneous reheating history \cite{Mizuno:2019bxy} (See also Refs. \cite{Dhuria:2019oyf,Torabian:2019zms} for nonthermal/nonstandard post-inflationary history). Starting with the observation that the current comoving Hubble horizon should be originated from the comoving Hubble horizon at the beginning of the inflation, $1/(a_0H_0)\lesssim1/(a_iH_i)$, one arrives at 
\begin{align}
\frac{H_i}{M_\mathrm{Pl}}<\frac{a_0H_0}{a_fH_f}
\end{align}
after appreciating the TCC bound [Eq. \eqref{eq:TCCa}]. To further evaluate the denominator $a_fH_f$ at the end of inflation in terms of a general reheating history characterized by an e-folding number $N_\mathrm{reh}$ and an equation-of-state (EOS) parameter $w_\mathrm{reh}$, one could use the following relations:
\begin{align}
&\frac{a_f}{a_\mathrm{reh}}=\mathrm{e}^{-N_\mathrm{reh}}, \quad \frac{a_\mathrm{reh}}{a_0}=\left(\frac{43}{11g_\mathrm{reh}}\right)^\frac13\frac{T_0}{T_\mathrm{reh}}, \label{eq:Treh1}\\
&3M_\mathrm{Pl}^2H_f^2\mathrm{e}^{-3N_\mathrm{reh}(1+w_\mathrm{reh})}=\frac{\pi^2}{30}g_\mathrm{reh}T_\mathrm{reh}^4,\label{eq:Treh2}
\end{align}
where $T_\mathrm{reh}$, $g_\mathrm{reh}$, and $a_\mathrm{reh}$ are the reheating temperature, the degrees of freedom of relativistic species, and the scale factor at the end of reheating, respectively. The inflationary Hubble scale is therefore bounded from above by 
\begin{align}\label{eq:TCCHnew}
\frac{H_i}{M_\mathrm{Pl}}&<\mathrm{e}^{-\frac12N_\mathrm{reh}(1+3w_\mathrm{reh})}\frac{(11/43)^{1/3}}{(\pi^2/90)^{1/2}}g_\mathrm{reh}^{-1/6}\frac{H_0M_\mathrm{Pl}}{T_0T_\mathrm{reh}}\\
&\lesssim\frac{H_0M_\mathrm{Pl}}{T_0T_\mathrm{reh}}\approx66\frac{T_0}{T_\mathrm{reh}}=1.5\times10^{-8}\left(\frac{1\,\mathrm{MeV}}{T_\mathrm{reh}}\right)
\end{align}
where the first inequality in the second line is taken for a near-critical expansion after inflation with the EOS parameter $w_\mathrm{reh}\gtrsim-1/3$ to achieve the maximum relaxation on the tensor-to-scalar ratio
\begin{align}\label{eq:TCCrnew}
r\lesssim2.3\times10^{-8}\left(\frac{1\,\mathrm{MeV}}{T_\mathrm{reh}}\right)^2.
\end{align}
Now the upper bound in Eq. \eqref{eq:TCCrnew} with the reheating temperature at the lowest possible temperature required by big bang nucleosynthesis (BBN) could be realized in some supergravity- or string-inspired inflation models. On the other hand, the reheating temperature cannot be too large; otherwise, the inflationary energy density bounded by Eq. \eqref{eq:TCCHnew} could be smaller than the reheating energy density. Therefore, by requiring $3M_\mathrm{Pl}^2H_i^2\gtrsim\frac{\pi^2}{30}g_\mathrm{reh}T_\mathrm{reh}^4$ for Eq. \eqref{eq:TCCHnew}, one obtains
\begin{align}
\left(\frac{T_\mathrm{reh}}{M_\mathrm{Pl}}\right)^6\lesssim\frac{90}{\pi^2}\frac{H_0^2}{T_0^2};
\end{align}
namely, $T_\mathrm{reh}\lesssim6.5\times10^8\,\mathrm{GeV}$. See Ref. \cite{Cai:2019hge,Tenkanen:2019wsd,Das:2019hto,Brahma:2019unn} for other discussions on the TCC from the viewpoints of initial state, dark matter, and  warm inflation.

In this paper, we will discuss the TCC implication on the  mass bound for the primordial black holes (PBHs) formed in the radiation-dominated era (Sec. \ref{subsec:RD}) and an early matter-dominated era (Sec. \ref{subsec:MD}), assuming that the PBH formation at reentry comes from large curvature perturbations at small scales. We conclude in the last section. It is worth noting that  the derived PBH mass bounds would not be applicable to other scenarios of PBH production from curvatons \cite{Kohri:2012yw,Kawasaki:2012wr}, scalar lumps  \cite{Cotner:2018vug,Cotner:2019ykd}, cosmic strings \cite{Hawking:1987bn,Polnarev:1988dh}, domain walls \cite{Deng:2016vzb,Liu:2019lul}, primordial bubbles \cite{Deng:2017uwc}, bubble collisions \cite{Kodama:1982sf,Hawking:1982ga}, and preheating instability \cite{Martin:2019nuw}, to name just a few. 

\section{Mass bound for PBH from TCC}\label{sec:bound}

The only existing lower bound on the PBH mass, $M_\mathrm{PBH}\gtrsim10^{15}$ g, comes from the observation of the absence of extragalactic photons \cite{Page:1976wx,Carr:2009jm} during PBH evaporation\cite{Hawking:1974rv,Hawking:1974sw}. Here we will derive the mass bound for PBH from a theoretical perspective of the TCC, irrespective of details of the reheating history.

\subsection{PBHs formed in the radiation-dominated era}\label{subsec:RD}

\begin{figure*}
\includegraphics[width=0.35\textwidth]{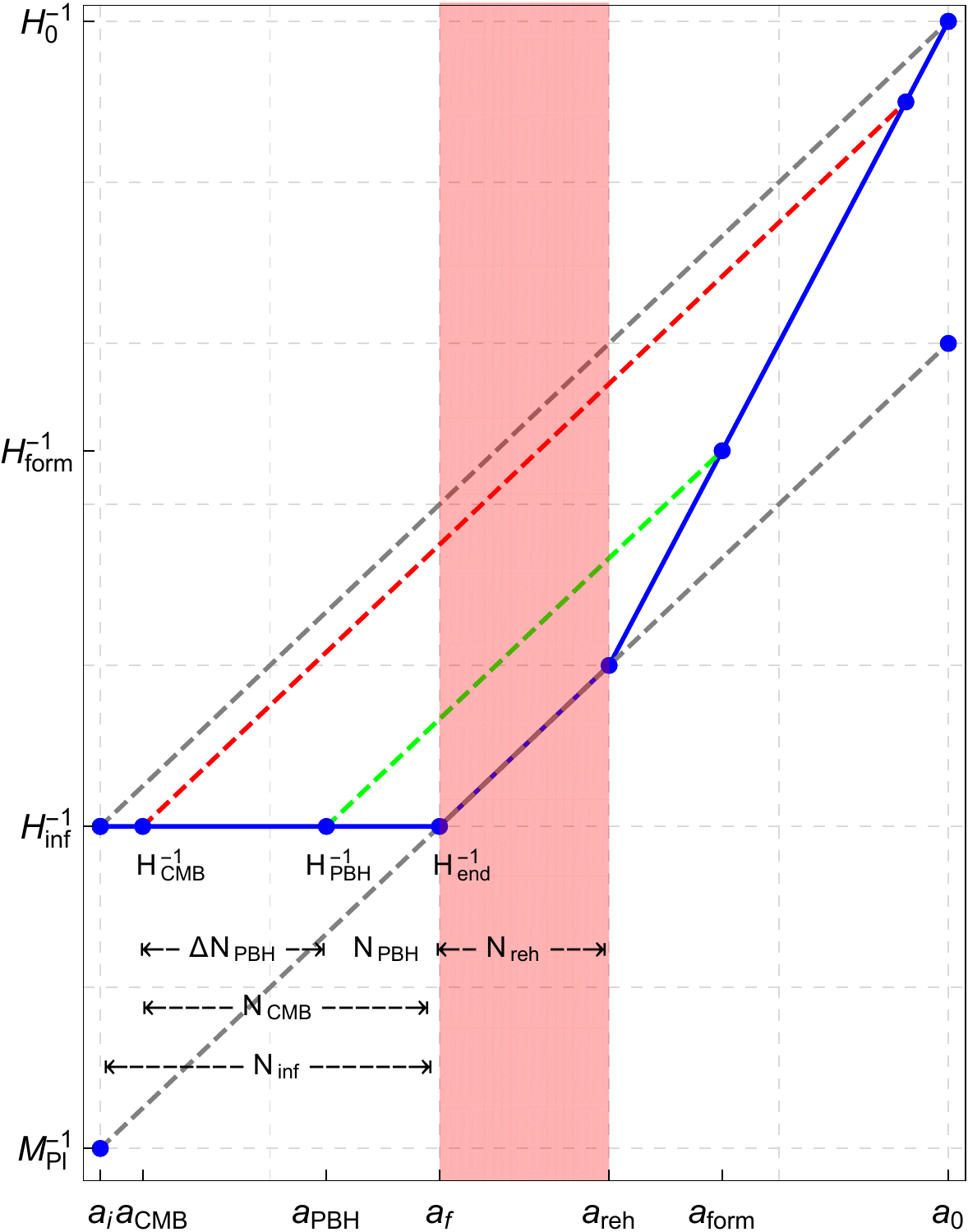}
\includegraphics[width=0.63\textwidth]{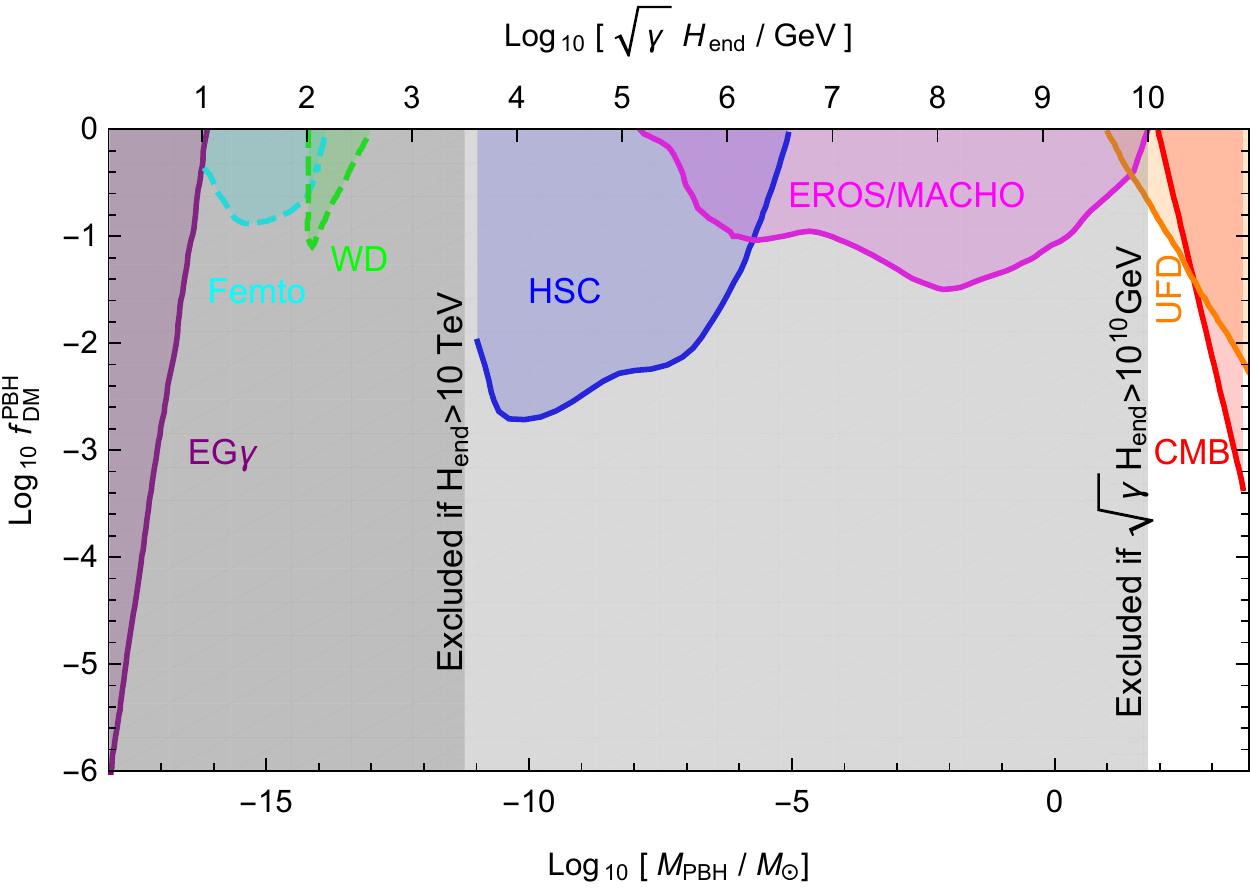}\\
\caption{\textit{Left}: Demonstration for PBH formation in the radiation-dominated era. The late matter- and dark-energy-dominated eras are omitted for clarity.  \textit{Right}: The implication of our lower bound on PBH mass. The colored regions are excluded by the observations ( EG$\gamma$: Extragalactic gamma ray \cite{Carr:2009jm}, Femto: Femtolensing \cite{Barnacka:2012bm}, WD: White dwarfs explosions \cite{Graham:2015apa}, HSC: Subaru Hyper Suprime-Cam  microlensing \cite{Niikura:2017zjd}, EROS/MACHO: EROS \cite{Tisserand:2006zx} and MACHO \cite{Allsman:2000kg}, UFD: Ultrafaint dwarfs \cite{Brandt:2016aco}, and CMB \cite{Ali-Haimoud:2016mbv,Poulin:2017bwe}) on the PBH abundance in DM for a given PBH mass. The gray region could be excluded by our lower bound on PBH mass from the TCC if the inflationary Hubble scale is larger than the certain value indicated by the top frame ticks.}\label{fig:PBHRD}
\end{figure*}

For PBHs collapsed from the horizon mass with efficiency factor $\gamma\approx0.2$ \cite{Carr:1975qj}, the PBH mass is estimated by
\begin{align}
M_\mathrm{PBH}=\gamma\frac43\pi H_\mathrm{form}^{-3}(3M_\mathrm{Pl}^2H_\mathrm{form}^2)=\frac{4\pi\gamma M_\mathrm{Pl}^2}{H_\mathrm{form}}.
\end{align}
For PBHs formed in the radiation-dominated era (as shown in the left panel of Fig. \ref{fig:PBHRD}), the Hubble scale at PBH formation, $H_\mathrm{form}$, should be at least smaller than the Hubble scale $H_\mathrm{end}$ at the end of inflation, which is further constrained by the TCC bound from the total e-folding number of inflation set by the current observable scale, namely
\begin{align}
M_\mathrm{PBH}
=\frac{4\pi\gamma M_\mathrm{Pl}^2}{H_\mathrm{form}}
\geq\frac{4\pi\gamma M_\mathrm{Pl}^2}{H_\mathrm{end}}
\geq4\pi\gamma M_\mathrm{Pl}\mathrm{e}^{N_\mathrm{inf}}.
\end{align}
Therefore, it is expected to see a lower bound on the PBH mass formed in the radiation era. 

To be specific, $H_\mathrm{form}$ could be related to the Hubble scale $H_\mathrm{PBH}$ at the exit of the corresponding curvature perturbations via the comoving relation $a_\mathrm{form}H_\mathrm{form}=a_\mathrm{PBH}H_\mathrm{PBH}$ and the scaling relation $H\propto a^{-3(1+w)/2}$, namely
\begin{align}
\frac{H_\mathrm{form}}{H_\mathrm{PBH}}
&=\frac{a_\mathrm{PBH}^2}{a_\mathrm{end}^2}\frac{a_\mathrm{end}^2}{a_\mathrm{reh}^2}\frac{a_\mathrm{reh}^2}{a_\mathrm{form}^2}\frac{H_\mathrm{PBH}}{H_\mathrm{form}}\nonumber\\
&=\mathrm{e}^{-2N_\mathrm{PBH}}\mathrm{e}^{-2N_\mathrm{reh}}\frac{H_\mathrm{form}}{H_\mathrm{reh}}\frac{H_\mathrm{PBH}}{H_\mathrm{form}}\nonumber\\
&=\mathrm{e}^{-2(N_\mathrm{PBH}+N_\mathrm{reh})}\frac{H_\mathrm{end}}{H_\mathrm{reh}}\frac{H_\mathrm{PBH}}{H_\mathrm{end}}\nonumber\\
&=\mathrm{e}^{-2(N_\mathrm{PBH}+N_\mathrm{reh})}\mathrm{e}^{\frac32N_\mathrm{reh}(1+w_\mathrm{reh})}\frac{H_\mathrm{PBH}}{H_\mathrm{end}}\nonumber\\
&=\mathrm{e}^{-2[N_\mathrm{PBH}+\frac14N_\mathrm{reh}(1-3w_\mathrm{reh})-\frac12\ln\frac{H_\mathrm{PBH}}{H_\mathrm{end}}]}.
\end{align}
The exponential factor in the above equation could be rearranged in such a way that all dependence on the reheating history could be totally removed by noting that \cite{Cai:2018rqf}
\begin{align}
&\frac{k_\mathrm{CMB}}{a_0H_0}
=\frac{a_\mathrm{CMB}}{a_\mathrm{end}}\frac{a_\mathrm{end}}{a_\mathrm{reh}}\frac{a_\mathrm{reh}}{a_0}\frac{H_\mathrm{CMB}}{H_0}\nonumber\\
&=\mathrm{e}^{-N_\mathrm{CMB}-N_\mathrm{reh}}\left(\frac{43}{11}\right)^{1/3}g_\mathrm{reh}^{-1/3}\frac{T_0}{T_\mathrm{reh}}\frac{H_\mathrm{CMB}}{H_0}\label{eq:Treh3}\\
&=\mathrm{e}^{-N_\mathrm{CMB}-\frac14N_\mathrm{reh}(1-3w_\mathrm{reh})}\frac{(\pi^2/90)^{1/4}}{(11/43)^{1/3}}g_\mathrm{reh}^{-1/12}\frac{T_0}{H_0}\frac{H_\mathrm{CMB}}{M_\mathrm{Pl}^{1/2}H_\mathrm{end}^{1/2}};\nonumber
\end{align}
namely,
\begin{align}
\frac14N_\mathrm{reh}(1-3w_\mathrm{reh})&=-N_\mathrm{CMB}+\frac12\ln\frac{H_\mathrm{CMB}^2}{M_\mathrm{Pl}H_\mathrm{end}}\nonumber\\
&+\ln\left[\frac{T_0}{k_\mathrm{CMB}}\frac{(\pi^2/90)^{1/4}}{(11/43)^{1/3}}g_\mathrm{reh}^{-1/12}\right],
\end{align}
where Eqs. \eqref{eq:Treh1} and \eqref{eq:Treh2} are used to replace $T_\mathrm{reh}$ in Eq.  \eqref{eq:Treh3}. Therefore, one finally arrives at
\begin{align}
\frac{H_\mathrm{form}}{H_\mathrm{PBH}}=\mathrm{e}^{-2[N_\mathrm{tot}-\Delta N_\mathrm{PBH}+\frac12\ln\frac{H_\mathrm{CMB}}{H_\mathrm{PBH}}]},
\end{align}
where $\Delta N_\mathrm{PBH}\equiv N_\mathrm{CMB}-N_\mathrm{PBH}$ is the difference in the e-folding number at the exit of the CMB pivot scale $k_\mathrm{CMB}=0.002\,\mathrm{Mpc}^{-1}$ with respect to the exit of curvature perturbations that collapse into PBHs at reentry, and $N_\mathrm{tot}$ is an abbreviation for the combination
\begin{align}
N_\mathrm{tot}&\equiv\ln\left[\frac{T_0}{k_\mathrm{CMB}}\frac{(\pi^2/90)^{1/4}}{(11/43)^{1/3}}g_\mathrm{reh}^{-1/12}\right]+\frac14\ln\left(\frac{\pi^2}{2}r\mathcal{P}_\mathcal{R}\right)\nonumber\\
&\approx64.99+\frac14\ln(r\mathcal{P}_\mathcal{R})\approx65+\frac14\ln\left(\frac{2}{\pi^2}\frac{H_\mathrm{CMB}^2}{M_\mathrm{Pl}^2}\right).
\end{align}
Now the PBH mass can be expressed as
\begin{align}
M_\mathrm{PBH}=4\sqrt{2}\gamma\left(\frac{H_\mathrm{CMB}}{H_\mathrm{PBH}}\right)^2\mathrm{e}^{2(65-\Delta N_\mathrm{PBH})}M_\mathrm{Pl}.
\end{align}
For PBH fluctuations reentering the horizon after the reheating era, $N_\mathrm{PBH}+N_\mathrm{reh}>\frac32N_\mathrm{reh}(1+w_\mathrm{reh})$, hence
\begin{align}
\Delta N_\mathrm{PBH}&\leq N_\mathrm{inf}-\frac12N_\mathrm{reh}(1+3w_\mathrm{reh})\nonumber\\
&\leq\ln\frac{M_\mathrm{Pl}}{H_\mathrm{end}}-\frac12N_\mathrm{reh}(1+3w_\mathrm{reh}),
\end{align}
where $N_\mathrm{CMB}<N_\mathrm{inf}$ and the TCC bound are used.
After using $H_\mathrm{CMB}\gtrsim H_\mathrm{PBH}$,  a lower bound is expected,
\begin{align}\label{eq:lowerbound}
M_\mathrm{PBH}&>\gamma\left(\frac{H_\mathrm{end}}{1.3\times10^9\,\mathrm{GeV}}\right)^2\mathrm{e}^{(1+3w_\mathrm{reh})N_\mathrm{reh}}\,M_\odot\nonumber\\
&>\gamma\left(\frac{H_\mathrm{end}}{1.3\times10^9\,\mathrm{GeV}}\right)^2\,M_\odot,
\end{align}
which is independent of specific configurations of reheating history. It is easy to see that if there is a lower bound on the inflationary Hubble scale, there would be a corresponding lower bound on the PBH mass, below which there are no PBHs, as shown by the excluded gray regions in the right panel of Fig. \ref{fig:PBHRD}. On the other hand, the PBH abundance in cold dark matter (DM) cannot be constrained by the TCC, since it is exponentially sensitive to the small-scale enhancement in curvature perturbations.

Several implications from our lower bound on PBH mass [Eq. \eqref{eq:lowerbound}] are as follows: First, the observational lower bound $M_\mathrm{PBH}\gtrsim10^{-18}\,M_\odot$ can always be fulfilled as long as the inflationary scale $H_\mathrm{end}\gtrsim\gamma^{-1/2}\,\mathrm{GeV}$. Second, no PBH with mass smaller than $10^2\,M_\odot$ is allowed if the inflationary Hubble scale is larger than $10^{10}\gamma^{-1/2}\,\mathrm{GeV}$. Fortunately, such LIGO-type PBHs could be allowed by the TCC, because the reheating-assisted TCC bound on  the inflationary Hubble scale [Eq. \eqref{eq:TCCHnew}] forbids any inflationary scale larger than $10^{10}\,\mathrm{GeV}$ unless the reheating temperature is lower than $1\,\mathrm{MeV}$ that would otherwise violate the BBN constraint on reheating temperature. Third, currently there is an open window \cite{Katz:2018zrn,Montero-Camacho:2019jte} below sublunar mass for PBHs making up all the cold DM; however, such a window could be totally closed if the inflationary Hubble scale is larger than 10 TeV scale.

\subsection{PBHs formed in an early matter-dominated era}\label{subsec:MD}

\begin{figure*}
\includegraphics[width=0.35\textwidth]{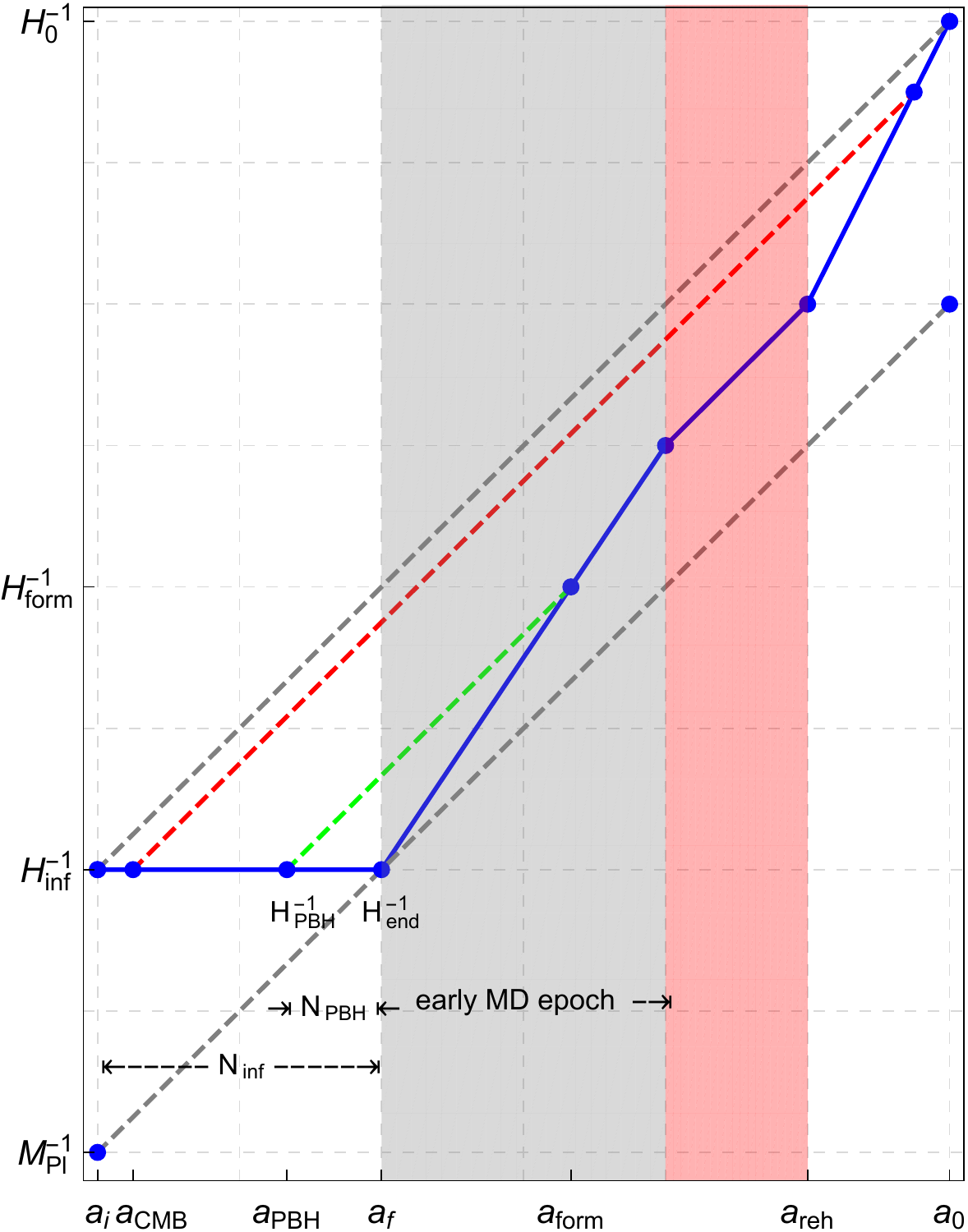}
\includegraphics[width=0.63\textwidth]{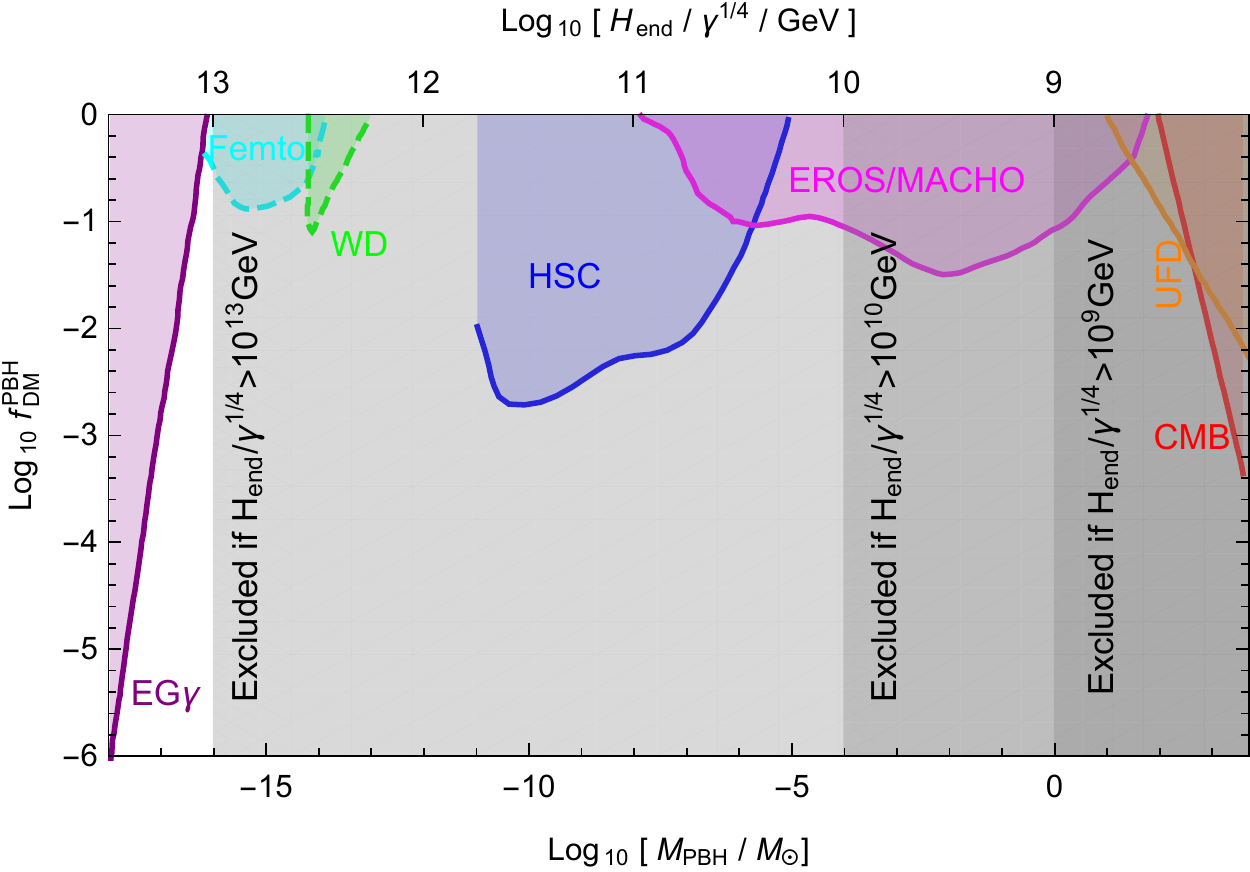}\\
\caption{\textit{Left}: Demonstration for PBH formation in an early matter-dominated era. \textit{Right}: The implication of our upper bound on PBH mass. The colored regions are excluded by the observations. The gray region could be excluded by our upper bound on PBH mass from TCC if the inflationary Hubble scale is larger than the certain value indicated by the top frame ticks.}\label{fig:PBHMD}
\end{figure*}

To minimally extend the previous discussion on mass bound for PBHs to other production channels, one could also consider PBH formation in an early matter-dominated era right after inflation but before the reheating era \cite{Harada:2016mhb,Carr:2017edp,Carr:2018nkm}, which is shown by the gray region in the left panel of Fig. \ref{fig:PBHMD}. After simple manipulations with the comoving relation $a_\mathrm{form}H_\mathrm{form}=a_\mathrm{PBH}H_\mathrm{PBH}$ and the scaling relation $H\propto a^{-3(1+w)/2}$, one could express the Hubble scale at PBH formation in terms of the Hubble scale at the exit of the enhanced small-scale fluctuations directly by 
\begin{align}
\frac{H_\mathrm{form}}{H_\mathrm{PBH}}&=\frac{a_\mathrm{PBH}}{a_\mathrm{form}}=\frac{a_\mathrm{PBH}^3}{a_\mathrm{end}^3}\frac{a_\mathrm{end}^3}{a_\mathrm{form}^3}\frac{H_\mathrm{PBH}^2}{H_\mathrm{form}^2}\nonumber\\
&=\mathrm{e}^{-3N_\mathrm{PBH}}\frac{H_\mathrm{PBH}^2}{H_\mathrm{end}^2}
\end{align}
without referring to the reheating era. Hence the PBH mass becomes
\begin{align}
M_\mathrm{PBH}&=4\pi\gamma\frac{M_\mathrm{Pl}^2H_\mathrm{end}^2}{H_\mathrm{PBH}^3}\mathrm{e}^{3N_\mathrm{PBH}}<4\pi\gamma\frac{M_\mathrm{Pl}^2H_\mathrm{end}^2}{H_\mathrm{PBH}^3}\mathrm{e}^{3N_\mathrm{inf}}\nonumber\\
&<4\pi\gamma\frac{M_\mathrm{Pl}^5}{H_\mathrm{PBH}^3H_\mathrm{end}}\lesssim4\pi\gamma\frac{M_\mathrm{Pl}^5}{H_\mathrm{end}^4},
\end{align}
where we have used $N_\mathrm{PBH}<N_\mathrm{inf}<\ln(M_\mathrm{Pl}/H_\mathrm{end})$ and $H_\mathrm{PBH}\gtrsim H_\mathrm{end}$. Now the PBH mass formed in an early matter-dominated era has an upper bound as
\begin{align}\label{eq:upperbound}
M_\mathrm{PBH}<\gamma\left(\frac{9.9\times10^8\,\mathrm{GeV}}{H_\mathrm{end}}\right)^4\,M_\odot,
\end{align}
above which is excluded, as shown in gray in the right panel of Fig. \ref{fig:PBHMD} if the inflationary scale is larger than a certain value. Several implications from the above bound are in order: First, no PBH with mass larger than $10^{-18}\,M_\odot$ is allowed if the inflationary Hubble scale is larger than $10^{14}\,\mathrm{GeV}$. Second, LIGO-type PBHs cannot fit the upper bound [Eq.  \eqref{eq:upperbound}] if the inflationary Hubble scale is larger than $10^9\,\mathrm{GeV}$. Third, PBHs formed in an early matter-dominated era with mass larger than $10^4\,M_\odot$ are not allowed if the inflationary Hubble scale is larger than $10^8\,\mathrm{GeV}$. 

\section{Conclusion}\label{sec:con}

In this paper, using only the recently proposed TCC bound on the inflationary e-folding number, we derive the mass bounds for PBHs formed in the radiation-dominated and early matter-dominated eras from the enhanced curvature perturbations at small scales in the inflationary cosmology. The explicit dependences on the detail configurations of reheating history are carefully removed. The resulting mass bounds for PBH therefore only rely on the inflationary Hubble scale. In particular, for PBHs formed in the radiation-dominated era, the asteroid-mass PBHs observationally allowed to make up all the cold DM cannot exist if the inflationary Hubble scale is higher than 10 TeV scale.

\begin{acknowledgments}
We thank Shi Pi for helpful discussion. R.-G. C. was supported by the National Natural Science Foundation of China Grants No. 11690022, No. 11821505 and No. 11851302, and by the Strategic Priority Research Program of Chinese Academy of Sciences Grant No. XDB23030100, and by the Key Research Program of Frontier Sciences of the Chinese Academy of Sciences. S.-J. W. is supported by the postdoctoral scholarship of Tufts University from the National Science Foundation.
\end{acknowledgments}

\bibliographystyle{utphys}
\bibliography{ref}

\providecommand{\href}[2]{#2}\begingroup\raggedright\begin{thebibliography}{10}

\bibitem{Gorski:1996cf}
K.~M. Gorski, A.~J. Banday, C.~L. Bennett, G.~Hinshaw, A.~Kogut, G.~F. Smoot,
  and E.~L. Wright, ``{Power spectrum of primordial inhomogeneity determined
  from the four year COBE DMR sky maps},''
  \href{http://dx.doi.org/10.1086/310077}{{\em Astrophys. J.} {\bfseries 464}
  (1996) L11},
\href{http://arxiv.org/abs/astro-ph/9601063}{{\ttfamily arXiv:astro-ph/9601063
  [astro-ph]}}.

\bibitem{Hinshaw:2012aka}
{\bfseries WMAP} Collaboration, G.~Hinshaw {\em et~al.}, ``{Nine-Year Wilkinson
  Microwave Anisotropy Probe (WMAP) Observations: Cosmological Parameter
  Results},'' \href{http://dx.doi.org/10.1088/0067-0049/208/2/19}{{\em
  Astrophys. J. Suppl.} {\bfseries 208} (2013) 19},
\href{http://arxiv.org/abs/1212.5226}{{\ttfamily arXiv:1212.5226
  [astro-ph.CO]}}.

\bibitem{Aghanim:2018eyx}
{\bfseries Planck} Collaboration, N.~Aghanim {\em et~al.}, ``{Planck 2018
  results. VI. Cosmological parameters},''
\href{http://arxiv.org/abs/1807.06209}{{\ttfamily arXiv:1807.06209
  [astro-ph.CO]}}.

\bibitem{Brout:1977ix}
R.~Brout, F.~Englert, and E.~Gunzig, ``{The Creation of the Universe as a
  Quantum Phenomenon},''
\href{http://dx.doi.org/10.1016/0003-4916(78)90176-8}{{\em Annals Phys.}
  {\bfseries 115} (1978) 78}.

\bibitem{Sato:1980yn}
K.~Sato, ``{First Order Phase Transition of a Vacuum and Expansion of the
  Universe},''
{\em Mon. Not. Roy. Astron. Soc.} {\bfseries 195} (1981) 467--479.

\bibitem{Guth:1980zm}
A.~H. Guth, ``{The Inflationary Universe: A Possible Solution to the Horizon
  and Flatness Problems},''
\href{http://dx.doi.org/10.1103/PhysRevD.23.347}{{\em Phys. Rev.} {\bfseries
  D23} (1981) 347--356}.

\bibitem{Linde:1981mu}
A.~D. Linde, ``{A New Inflationary Universe Scenario: A Possible Solution of
  the Horizon, Flatness, Homogeneity, Isotropy and Primordial Monopole
  Problems},''
\href{http://dx.doi.org/10.1016/0370-2693(82)91219-9}{{\em Phys. Lett.}
  {\bfseries B108} (1982) 389--393}.

\bibitem{Albrecht:1982wi}
A.~Albrecht and P.~J. Steinhardt, ``{Cosmology for Grand Unified Theories with
  Radiatively Induced Symmetry Breaking},''
\href{http://dx.doi.org/10.1103/PhysRevLett.48.1220}{{\em Phys. Rev. Lett.}
  {\bfseries 48} (1982) 1220--1223}.

\bibitem{Brandenberger:2009jq}
R.~H. Brandenberger, ``{Alternatives to the inflationary paradigm of structure
  formation},'' \href{http://dx.doi.org/10.1142/S2010194511000109}{{\em Int. J.
  Mod. Phys. Conf. Ser.} {\bfseries 01} (2011) 67--79},
\href{http://arxiv.org/abs/0902.4731}{{\ttfamily arXiv:0902.4731 [hep-th]}}.

\bibitem{Martin:2000xs}
J.~Martin and R.~H. Brandenberger, ``{The TransPlanckian problem of
  inflationary cosmology},''
  \href{http://dx.doi.org/10.1103/PhysRevD.63.123501}{{\em Phys. Rev.}
  {\bfseries D63} (2001) 123501},
\href{http://arxiv.org/abs/hep-th/0005209}{{\ttfamily arXiv:hep-th/0005209
  [hep-th]}}.

\bibitem{Brandenberger:2000wr}
R.~H. Brandenberger and J.~Martin, ``{The Robustness of inflation to changes in
  superPlanck scale physics},''
  \href{http://dx.doi.org/10.1142/S0217732301004170}{{\em Mod. Phys. Lett.}
  {\bfseries A16} (2001) 999--1006},
\href{http://arxiv.org/abs/astro-ph/0005432}{{\ttfamily arXiv:astro-ph/0005432
  [astro-ph]}}.

\bibitem{Brandenberger:2012aj}
R.~H. Brandenberger and J.~Martin, ``{Trans-Planckian Issues for Inflationary
  Cosmology},'' \href{http://dx.doi.org/10.1088/0264-9381/30/11/113001}{{\em
  Class. Quant. Grav.} {\bfseries 30} (2013) 113001},
\href{http://arxiv.org/abs/1211.6753}{{\ttfamily arXiv:1211.6753
  [astro-ph.CO]}}.

\bibitem{Kaloper:2002cs}
N.~Kaloper, M.~Kleban, A.~Lawrence, S.~Shenker, and L.~Susskind, ``{Initial
  conditions for inflation},''
  \href{http://dx.doi.org/10.1088/1126-6708/2002/11/037}{{\em JHEP} {\bfseries
  11} (2002) 037},
\href{http://arxiv.org/abs/hep-th/0209231}{{\ttfamily arXiv:hep-th/0209231
  [hep-th]}}.

\bibitem{Easther:2002xe}
R.~Easther, B.~R. Greene, W.~H. Kinney, and G.~Shiu, ``{A Generic estimate of
  transPlanckian modifications to the primordial power spectrum in
  inflation},'' \href{http://dx.doi.org/10.1103/PhysRevD.66.023518}{{\em Phys.
  Rev.} {\bfseries D66} (2002) 023518},
\href{http://arxiv.org/abs/hep-th/0204129}{{\ttfamily arXiv:hep-th/0204129
  [hep-th]}}.

\bibitem{Bedroya:2019snp}
A.~Bedroya and C.~Vafa, ``{Trans-Planckian Censorship and the Swampland},''
\href{http://arxiv.org/abs/1909.11063}{{\ttfamily arXiv:1909.11063 [hep-th]}}.

\bibitem{ArkaniHamed:2007ky}
N.~Arkani-Hamed, S.~Dubovsky, A.~Nicolis, E.~Trincherini, and G.~Villadoro,
  ``{A Measure of de Sitter entropy and eternal inflation},''
  \href{http://dx.doi.org/10.1088/1126-6708/2007/05/055}{{\em JHEP} {\bfseries
  05} (2007) 055},
\href{http://arxiv.org/abs/0704.1814}{{\ttfamily arXiv:0704.1814 [hep-th]}}.

\bibitem{Bedroya:2019tba}
A.~Bedroya, R.~Brandenberger, M.~Loverde, and C.~Vafa, ``{Trans-Planckian
  Censorship and Inflationary Cosmology},''
\href{http://arxiv.org/abs/1909.11106}{{\ttfamily arXiv:1909.11106 [hep-th]}}.

\bibitem{Mizuno:2019bxy}
S.~Mizuno, S.~Mukohyama, S.~Pi, and Y.-L. Zhang, ``{Universal Upper Bound on
  the Inflationary Energy Scale from the Trans-Planckian Censorship
  Conjecture},''
\href{http://arxiv.org/abs/1910.02979}{{\ttfamily arXiv:1910.02979
  [astro-ph.CO]}}.

\bibitem{Dhuria:2019oyf}
M.~Dhuria and G.~Goswami, ``{Trans-Planckian Censorship Conjecture and
  Non-thermal post-inflationary history},''
\href{http://arxiv.org/abs/1910.06233}{{\ttfamily arXiv:1910.06233
  [astro-ph.CO]}}.

\bibitem{Torabian:2019zms}
M.~Torabian, ``{Non-Standard Cosmological Models and the trans-Planckian
  Censorship Conjecture},''
\href{http://arxiv.org/abs/1910.06867}{{\ttfamily arXiv:1910.06867 [hep-th]}}.

\bibitem{Cai:2019hge}
Y.~Cai and Y.-S. Piao, ``{Pre-inflation and Trans-Planckian Censorship},''
\href{http://arxiv.org/abs/1909.12719}{{\ttfamily arXiv:1909.12719 [gr-qc]}}.

\bibitem{Tenkanen:2019wsd}
T.~Tenkanen, ``{Trans-Planckian Censorship, Inflation and Dark Matter},''
\href{http://arxiv.org/abs/1910.00521}{{\ttfamily arXiv:1910.00521
  [astro-ph.CO]}}.

\bibitem{Das:2019hto}
S.~Das, ``{Distance, de Sitter and Trans-Planckian Censorship conjectures: the
  status quo of Warm Inflation},''
\href{http://arxiv.org/abs/1910.02147}{{\ttfamily arXiv:1910.02147 [hep-th]}}.

\bibitem{Brahma:2019unn}
S.~Brahma, ``{Trans-Planckian censorship, inflation and excited initial states
  for perturbations},''
\href{http://arxiv.org/abs/1910.04741}{{\ttfamily arXiv:1910.04741 [hep-th]}}.

\bibitem{Kohri:2012yw}
K.~Kohri, C.-M. Lin, and T.~Matsuda, ``{Primordial black holes from the
  inflating curvaton},''
  \href{http://dx.doi.org/10.1103/PhysRevD.87.103527}{{\em Phys. Rev.}
  {\bfseries D87} no.~10, (2013) 103527},
\href{http://arxiv.org/abs/1211.2371}{{\ttfamily arXiv:1211.2371 [hep-ph]}}.

\bibitem{Kawasaki:2012wr}
M.~Kawasaki, N.~Kitajima, and T.~T. Yanagida, ``{Primordial black hole
  formation from an axionlike curvaton model},''
  \href{http://dx.doi.org/10.1103/PhysRevD.87.063519}{{\em Phys. Rev.}
  {\bfseries D87} no.~6, (2013) 063519},
\href{http://arxiv.org/abs/1207.2550}{{\ttfamily arXiv:1207.2550 [hep-ph]}}.

\bibitem{Cotner:2018vug}
E.~Cotner, A.~Kusenko, and V.~Takhistov, ``{Primordial Black Holes from
  Inflaton Fragmentation into Oscillons},''
  \href{http://dx.doi.org/10.1103/PhysRevD.98.083513}{{\em Phys. Rev.}
  {\bfseries D98} no.~8, (2018) 083513},
\href{http://arxiv.org/abs/1801.03321}{{\ttfamily arXiv:1801.03321
  [astro-ph.CO]}}.

\bibitem{Cotner:2019ykd}
E.~Cotner, A.~Kusenko, M.~Sasaki, and V.~Takhistov, ``{Analytic Description of
  Primordial Black Hole Formation from Scalar Field Fragmentation},''
\href{http://arxiv.org/abs/1907.10613}{{\ttfamily arXiv:1907.10613
  [astro-ph.CO]}}.

\bibitem{Hawking:1987bn}
S.~W. Hawking, ``{Black Holes From Cosmic Strings},''
\href{http://dx.doi.org/10.1016/0370-2693(89)90206-2}{{\em Phys. Lett.}
  {\bfseries B231} (1989) 237--239}.

\bibitem{Polnarev:1988dh}
A.~Polnarev and R.~Zembowicz, ``{Formation of Primordial Black Holes by Cosmic
  Strings},''
\href{http://dx.doi.org/10.1103/PhysRevD.43.1106}{{\em Phys. Rev.} {\bfseries
  D43} (1991) 1106--1109}.

\bibitem{Deng:2016vzb}
H.~Deng, J.~Garriga, and A.~Vilenkin, ``{Primordial black hole and wormhole
  formation by domain walls},''
  \href{http://dx.doi.org/10.1088/1475-7516/2017/04/050}{{\em JCAP} {\bfseries
  1704} no.~04, (2017) 050},
\href{http://arxiv.org/abs/1612.03753}{{\ttfamily arXiv:1612.03753 [gr-qc]}}.

\bibitem{Liu:2019lul}
J.~Liu, Z.-K. Guo, and R.-G. Cai, ``{Primordial Black Holes from Cosmic Domain
  Walls},''
\href{http://arxiv.org/abs/1908.02662}{{\ttfamily arXiv:1908.02662
  [astro-ph.CO]}}.

\bibitem{Deng:2017uwc}
H.~Deng and A.~Vilenkin, ``{Primordial black hole formation by vacuum
  bubbles},'' \href{http://dx.doi.org/10.1088/1475-7516/2017/12/044}{{\em JCAP}
  {\bfseries 1712} no.~12, (2017) 044},
\href{http://arxiv.org/abs/1710.02865}{{\ttfamily arXiv:1710.02865 [gr-qc]}}.

\bibitem{Kodama:1982sf}
H.~Kodama, M.~Sasaki, and K.~Sato, ``{Abundance of Primordial Holes Produced by
  Cosmological First Order Phase Transition},''
\href{http://dx.doi.org/10.1143/PTP.68.1979}{{\em Prog. Theor. Phys.}
  {\bfseries 68} (1982) 1979}.

\bibitem{Hawking:1982ga}
S.~W. Hawking, I.~G. Moss, and J.~M. Stewart, ``{Bubble Collisions in the Very
  Early Universe},''
\href{http://dx.doi.org/10.1103/PhysRevD.26.2681}{{\em Phys. Rev.} {\bfseries
  D26} (1982) 2681}.

\bibitem{Martin:2019nuw}
J.~Martin, T.~Papanikolaou, and V.~Vennin, ``{Primordial black holes from the
  preheating instability},''
\href{http://arxiv.org/abs/1907.04236}{{\ttfamily arXiv:1907.04236
  [astro-ph.CO]}}.

\bibitem{Page:1976wx}
D.~N. Page and S.~W. Hawking, ``{Gamma rays from primordial black holes},''
\href{http://dx.doi.org/10.1086/154350}{{\em Astrophys. J.} {\bfseries 206}
  (1976) 1--7}.

\bibitem{Carr:2009jm}
B.~J. Carr, K.~Kohri, Y.~Sendouda, and J.~Yokoyama, ``{New cosmological
  constraints on primordial black holes},''
  \href{http://dx.doi.org/10.1103/PhysRevD.81.104019}{{\em Phys. Rev.}
  {\bfseries D81} (2010) 104019},
\href{http://arxiv.org/abs/0912.5297}{{\ttfamily arXiv:0912.5297
  [astro-ph.CO]}}.

\bibitem{Hawking:1974rv}
S.~W. Hawking, ``{Black hole explosions},''
\href{http://dx.doi.org/10.1038/248030a0}{{\em Nature} {\bfseries 248} (1974)
  30--31}.

\bibitem{Hawking:1974sw}
S.~W. Hawking, ``{Particle Creation by Black Holes},''
  \href{http://dx.doi.org/10.1007/BF02345020, 10.1007/BF01608497}{{\em Commun.
  Math. Phys.} {\bfseries 43} (1975) 199--220}.
[,167(1975)].

\bibitem{Barnacka:2012bm}
A.~Barnacka, J.~F. Glicenstein, and R.~Moderski, ``{New constraints on
  primordial black holes abundance from femtolensing of gamma-ray bursts},''
  \href{http://dx.doi.org/10.1103/PhysRevD.86.043001}{{\em Phys. Rev.}
  {\bfseries D86} (2012) 043001},
\href{http://arxiv.org/abs/1204.2056}{{\ttfamily arXiv:1204.2056
  [astro-ph.CO]}}.

\bibitem{Graham:2015apa}
P.~W. Graham, S.~Rajendran, and J.~Varela, ``{Dark Matter Triggers of
  Supernovae},'' \href{http://dx.doi.org/10.1103/PhysRevD.92.063007}{{\em Phys.
  Rev.} {\bfseries D92} no.~6, (2015) 063007},
\href{http://arxiv.org/abs/1505.04444}{{\ttfamily arXiv:1505.04444 [hep-ph]}}.

\bibitem{Niikura:2017zjd}
H.~Niikura, M.~Takada, N.~Yasuda, R.~H. Lupton, T.~Sumi, S.~More, A.~More,
  M.~Oguri, and M.~Chiba, ``{Microlensing constraints on primordial black holes
  with the Subaru/HSC Andromeda observation},''
\href{http://arxiv.org/abs/1701.02151}{{\ttfamily arXiv:1701.02151
  [astro-ph.CO]}}.

\bibitem{Tisserand:2006zx}
{\bfseries EROS-2} Collaboration, P.~Tisserand {\em et~al.}, ``{Limits on the
  Macho Content of the Galactic Halo from the EROS-2 Survey of the Magellanic
  Clouds},'' \href{http://dx.doi.org/10.1051/0004-6361:20066017}{{\em Astron.
  Astrophys.} {\bfseries 469} (2007) 387--404},
\href{http://arxiv.org/abs/astro-ph/0607207}{{\ttfamily arXiv:astro-ph/0607207
  [astro-ph]}}.

\bibitem{Allsman:2000kg}
{\bfseries Macho} Collaboration, R.~A. Allsman {\em et~al.}, ``{MACHO project
  limits on black hole dark matter in the 1-30 solar mass range},''
  \href{http://dx.doi.org/10.1086/319636}{{\em Astrophys. J.} {\bfseries 550}
  (2001) L169},
\href{http://arxiv.org/abs/astro-ph/0011506}{{\ttfamily arXiv:astro-ph/0011506
  [astro-ph]}}.

\bibitem{Brandt:2016aco}
T.~D. Brandt, ``{Constraints on MACHO Dark Matter from Compact Stellar Systems
  in Ultra-Faint Dwarf Galaxies},''
  \href{http://dx.doi.org/10.3847/2041-8205/824/2/L31}{{\em Astrophys. J.}
  {\bfseries 824} no.~2, (2016) L31},
\href{http://arxiv.org/abs/1605.03665}{{\ttfamily arXiv:1605.03665
  [astro-ph.GA]}}.

\bibitem{Ali-Haimoud:2016mbv}
Y.~Ali-Haïmoud and M.~Kamionkowski, ``{Cosmic microwave background limits on
  accreting primordial black holes},''
  \href{http://dx.doi.org/10.1103/PhysRevD.95.043534}{{\em Phys. Rev.}
  {\bfseries D95} no.~4, (2017) 043534},
\href{http://arxiv.org/abs/1612.05644}{{\ttfamily arXiv:1612.05644
  [astro-ph.CO]}}.

\bibitem{Poulin:2017bwe}
V.~Poulin, P.~D. Serpico, F.~Calore, S.~Clesse, and K.~Kohri, ``{CMB bounds on
  disk-accreting massive primordial black holes},''
  \href{http://dx.doi.org/10.1103/PhysRevD.96.083524}{{\em Phys. Rev.}
  {\bfseries D96} no.~8, (2017) 083524},
\href{http://arxiv.org/abs/1707.04206}{{\ttfamily arXiv:1707.04206
  [astro-ph.CO]}}.

\bibitem{Carr:1975qj}
B.~J. Carr, ``{The Primordial black hole mass spectrum},''
\href{http://dx.doi.org/10.1086/153853}{{\em Astrophys. J.} {\bfseries 201}
  (1975) 1--19}.

\bibitem{Cai:2018rqf}
R.-G. Cai, T.-B. Liu, and S.-J. Wang, ``{Sensitivity of primordial black hole
  abundance on the reheating phase},''
  \href{http://dx.doi.org/10.1103/PhysRevD.98.043538}{{\em Phys. Rev.}
  {\bfseries D98} no.~4, (2018) 043538},
\href{http://arxiv.org/abs/1806.05390}{{\ttfamily arXiv:1806.05390
  [astro-ph.CO]}}.

\bibitem{Katz:2018zrn}
A.~Katz, J.~Kopp, S.~Sibiryakov, and W.~Xue, ``{Femtolensing by Dark Matter
  Revisited},'' \href{http://dx.doi.org/10.1088/1475-7516/2018/12/005}{{\em
  JCAP} {\bfseries 1812} (2018) 005},
\href{http://arxiv.org/abs/1807.11495}{{\ttfamily arXiv:1807.11495
  [astro-ph.CO]}}.

\bibitem{Montero-Camacho:2019jte}
P.~Montero-Camacho, X.~Fang, G.~Vasquez, M.~Silva, and C.~M. Hirata,
  ``{Revisiting constraints on asteroid-mass primordial black holes as dark
  matter candidates},''
\href{http://arxiv.org/abs/1906.05950}{{\ttfamily arXiv:1906.05950
  [astro-ph.CO]}}.

\bibitem{Harada:2016mhb}
T.~Harada, C.-M. Yoo, K.~Kohri, K.-i. Nakao, and S.~Jhingan, ``{Primordial
  black hole formation in the matter-dominated phase of the Universe},''
  \href{http://dx.doi.org/10.3847/1538-4357/833/1/61}{{\em Astrophys. J.}
  {\bfseries 833} no.~1, (2016) 61},
\href{http://arxiv.org/abs/1609.01588}{{\ttfamily arXiv:1609.01588
  [astro-ph.CO]}}.

\bibitem{Carr:2017edp}
B.~Carr, T.~Tenkanen, and V.~Vaskonen, ``{Primordial black holes from inflaton
  and spectator field perturbations in a matter-dominated era},''
  \href{http://dx.doi.org/10.1103/PhysRevD.96.063507}{{\em Phys. Rev.}
  {\bfseries D96} no.~6, (2017) 063507},
\href{http://arxiv.org/abs/1706.03746}{{\ttfamily arXiv:1706.03746
  [astro-ph.CO]}}.

\bibitem{Carr:2018nkm}
B.~Carr, K.~Dimopoulos, C.~Owen, and T.~Tenkanen, ``{Primordial Black Hole
  Formation During Slow Reheating After Inflation},''
\href{http://arxiv.org/abs/1804.08639}{{\ttfamily arXiv:1804.08639
  [astro-ph.CO]}}.

\end{thebibliography}\endgroup

\end{document}